\documentclass[%
 reprint,
 amsmath,amssymb,
 aps,
]{revtex4-2}

\usepackage{graphicx}
\usepackage{dcolumn}
\usepackage{bm}

\begin{document}

\preprint{APS/123-QED}

\title{Self-contained Beta-with-Spikes Approximation \\ for Inference Under a Wright-Fisher Model}

\author{Juan \surname{Guerrero Montero}}
    \email{J.A.Guererro-
Montero@sms.ed.ac.uk}
\author{Richard A. Blythe$^{†}$}%
\affiliation{%
 SUPA, School of Physics and Astronomy, University of Edinburgh, Edinburgh, EH9 3FD, United Kingdom 
}

\date{\today}

\begin{abstract}
We construct a reliable estimation method for evolutionary parameters within the Wright-Fisher model, which describes changes in allele frequencies due to selection and genetic drift, from time-series data. Such data exist for biological populations, for example via artificial evolution experiments, and for the cultural evolution of behavior, such as linguistic corpora that document historical usage of different words with similar meanings. Our method of analysis builds on a Beta-with-Spikes approximation to the distribution of allele frequencies predicted by the Wright-Fisher model. We introduce a self-contained scheme for estimating parameters in the approximation, and demonstrate its robustness with synthetic data, especially in the strong-selection and near-extinction regimes where previous approaches fail. We further apply the method to allele frequency data for baker’s yeast (\textit{Saccharomyces cerevisiae}), finding a significant signal of selection in cases where independent evidence supports such a conclusion. We further demonstrate the possibility of detecting time-points at which evolutionary parameters change in the context of a historical spelling reform in the Spanish language.
\end{abstract}

\maketitle

\section{Introduction}

Stochastic models have a long history in population genetics as a tool to understand the fate of populations undergoing evolution and to draw inferences about the demographic and evolutionary forces that have shaped a present-day population. The starting point of many analyses is the Wright-Fisher model \citep{ref:Wright,ref:Fisher,ref:WFBook} which characterizes the fluctuations that naturally occur as genetic material is replicated (genetic drift) and how these interact with mutation and selection. One key question is the extent to which variation in a population can be explained by a neutral model \citep{ref:Kimura1983}, that is one that appeals to genetic drift operating in the absence of selection. To this end a variety of classical statistical tests were developed \citep[see e.g.][for a review]{ref:Kreitmann2000} to detect departure from predictions of the neutral theory. Traditionally these are based on quantities that can be ascertained from a sample of genetic material taken from a population at a single time, such as the number of nucleotide differences.

Latterly, interest has turned to the analysis of time-series data, in particular measurements of the frequency of alleles in a population at multiple points in time. Historically, such information has been obtained from evolution experiments conducted in the laboratory \citep{ref:Lenski1991}. Meanwhile, advances in high-throughput sequencing technologies admit the collection of large datasets \citep{ref:Reuter2015}, and there is particular interest  in sampling microbial populations at multiple points in time \citep{ref:DominguezBello2011}. In addition to the expectation that additional time points will boost inferential power, the ability to analyze time series further opens the door to applications to other evolutionary paradigms, for example cultural evolution \citep{ref:BoydRicherson,ref:Cavalli}, in which analogs of nucleotide differences might not exist, but counterparts to allele frequencies do. The evolutionary process of language change is a case where one can identify processes that parallel mutation and selection \citep{ref:CroftLanguageChange}, which can furthermore be represented by a Wright-Fisher model \citep{ref:UtteranceSelection,ref:WordsAsAlleles,ref:RichardPLOS} and where large-scale historical records of variation are available \citep{ref:GoogleBooks,ref:COHA}. This broader context motivates further questions, such as whether one can detect changes in the sociocultural environment---for example, shifting attitudes towards specific behaviors---through changes in the intensity of evolutionary forces over time.

In this work, we develop a method to estimate the effective population size and selection strength in the Wright-Fisher model, to include the possibility that they each change over time, from a set of allele (or variant) frequencies obtained at different points in time. As in other works with a similar aim \citep[see][for reviews]{ref:Tataru2,ref:BwSComparisonTechnicalParis}, the basic strategy is to construct the likelihood of the observed parameter values and maximize with respect to the model parameters. This likelihood function involves the distribution of allele frequencies (DAF), conditioned on some initial value. Its construction has proved challenging: where exact solutions for the Wright-Fisher model exist, they are restricted to specific regimes and are cumbersome to work with, comprising for example an infinite series over special functions in the case of the diffusion approximation to the neutral Wright-Fisher model \citep{ref:WFBook}.

These difficulties have motivated a variety of schemes for approximating the DAF \citep{ref:Tataru2,ref:BwSComparisonTechnicalParis}. First, one can numerically integrate the diffusion equation corresponding to model of interest \citep{ref:bollback2008estimation}. Although still computationally demanding, this has the advantage of being able to incorporate arbitrary evolutionary forces, such as frequency-dependent selection. It is possible to reduce these demands by expressing the DAF in terms of a sequence of orthogonal polynomials and truncating at a low order \citep{ref:Lukic2012}, although this requires some care in dealing with artifacts arising from the truncation. Another way to proceed is to approximate the DAF with an appropriate distribution function, such as a Gaussian \citep{ref:lacerda2014population} or a Beta distribution \citep{ref:Hui2015}, and fix parameters by matching its moments to those obtained from the Wright-Fisher model. Such distribution functions are well-behaved by construction, but may fail to adequately approximate the true DAF \citep{ref:BwSComparisonTechnicalParis}. A separate line of attack is offered by the coalescent process \citep{ref:Kingman1982}, which is  dual to the Wright-Fisher model and particularly well-adapted to reconstructing genealogies \citep{ref:Siren2011}. However, such methods lend themselves most naturally to neutral evolution, and become somewhat more complex in the presence of selection.

Here, we pursue the approach of approximating the DAF with a distribution function that is sufficiently rich to capture the properties of the underlying Wright-Fisher model, but has a small number of parameters that can be estimated efficiently. Specifically, we adopt the Beta-with-Spikes (BwS) distribution introduced by \citet{ref:BWS1} as the functional form, and introduce a self-contained scheme that iteratively generates parameter values over multiple generations. The Beta distribution has been found to better describe changes in allele frequencies than a Gaussian distribution \citep{ref:BwSComparisonTechnicalParis}, primarily because the requirement that allele frequencies lie between $0$ and $1$ means that frequency differences are necessarily non-Gaussian as these boundary points are approached. Replacing the Gaussian with a Beta distribution rectifies this problem, but fails to account adequately for the accumulation of probability at the boundaries as individual realizations of the evolutionary dynamics cause an allele to reach fixation. It is precisely this shortcoming that augmenting the Beta distribution with delta functions (spikes) at the boundary points seeks to address \citep{ref:BWS1}. By estimating fixation probabilities and moments of the Wright-Fisher DAF, the parameters in the BwS distribution can then be chosen to match: this is then found to improve the approximation to the exact DAF relative to the Gaussian or standard Beta approximations \citep{ref:BwSComparisonTechnicalParis}.

In \cite{ref:Tataru2}, the moment-matching procedure is based on recursion relations for the mean and variance of the Wright-Fisher DAF that are not exact: they are based on a Taylor series expansion of the fitness function around the mean allele frequency in the previous generation. This approximation breaks down when selection is large \citep{ref:lacerda2014population}, and errors accumulate over multiple generations, sometimes to the point of exiting the parameter regime for which the BwS distribution is well-defined. The self-contained approach to estimating the BwS parameters that we introduce here avoids these problems. The basic idea is to start with a BwS distribution at the beginning of one generation, and to determine the parameter values that best approximate the distribution that results after one generation of evolution within the Wright-Fisher model. This leads to a \emph{self-contained} recursion, in the sense that we map the BwS parameters directly from one generation to the next via averages with respect to the BwS distribution, rather than via the moments of the target Wright-Fisher distribution. 

\section{Methods}

In this Section, we set out the method for obtaining the self-contained approximation for parameters in the Beta-with-Spikes distribution, and use synthetic data to compare the quality of the approximation with that obtained with previous moment-based approaches. We also outline how to use this approximation to obtain maximum-likelihood estimates for the effective population size and selection coefficient in the Wright-Fisher model, including the case where these change over time. In the following Section we will apply these methods to both genetic and cultural data sets.

\subsection{Wright-Fisher model}

The Wright-Fisher model describes the evolution of a population of randomly replicating individuals belonging to two or more types. In genetics, these types correspond to different alleles; in cultural evolution to variant forms of some socially-learned behavior. We consider here the case of two variants, denoted $A$ and $a$, with $x_{t}$ representing the proportion of $A$ individuals at generation $t$. The total population size, $N$, is assumed fixed, and each of the individuals in generation $t+1$ has probability $g(x_t)$ of being assigned type $A$ (otherwise they are assigned type $a$). We refer to $g(x)$ as the \emph{fitness function}, and it can be interpreted as the mean number of offspring that each $A$ leaves in the next generation. Given the above, the probability that the proportion of type $A$ individuals in generation $t+1$ is equal to $x_{t+1}$ is given by the binomial distribution
\begin{equation}\label{eq:WFTransitionProb}
    {\rm Pr}_{\rm WF}(x_{t+1}|x_{t})=
    \binom{N}{Nx_{t+1}}
    g\left(x_{t}\right)^{Nx_{t+1}}\left(1-g\left(x_{t}\right)\right)^{N(1-x_{t+1})} \;.
\end{equation}

In population genetics, it is well understood that structured populations, where individuals are divided into classes by age, sex, location or other characteristics, can be approximated by a Wright-Fisher model by setting $N$ equal to an appropriate \emph{effective} population size \citep{ref:Charlesworth2009}. The interpretation of $N$ is less obvious in the cultural case; however concrete models of language use have found $N$ to depend on factors like the size and structure of the speech community, the memory lifetime of individual speakers and the number of tokens produced in an utterance \citep{ref:UtteranceSelection,ref:WordsAsAlleles,ref:RichardPLOS}.  In any case, $N$ quantifies the effects of drift in the transmission: the lower $N$, the higher the uncertainty in the transmission to the next generation.

The form of $g(x)$ can be chosen to incorporate a variety of evolutionary forces, such as mutation, migration and selection of different types. Here we focus on the case of frequency-independent selection, which is traditionally modeled by assigning a weight (relative reproductive success) of $1+s_0$ to $A$ and $1$ to $a$. This yields a fitness function $\frac{(1+s_0)x}{1+s_0x}$. Here, we have found it helpful to adopt a different parametrization, in which the weights are $e^{s/2}$ and $e^{-s/2}$, respectively. This formulation has two particularly appealing features.

First, changing the sign of $s$ is equivalent to exchanging $A$ and $a$: that is, there is a symmetry between positive and negative values of $s$. This further implies that while we need $s_0\ge-1$ for the model to be well-defined, $s$ can take any positive or negative value. The second feature is that the fitness function
\begin{equation}\label{eq:s}
    g(x;s)=\frac{x}{x+(1-x)e^{-s}}
\end{equation}
satisfies the functional relation $g(g(x;s_1),s_2)=g(x;s_1+s_2)$. Thus in the absence of fluctuations, the evolution over $k$ generations with selection coefficients $s_1, s_2, \ldots, s_k$ is the same as a single generation of evolution with selection coefficient $s_1+s_2+\cdots+s_k$. Below, we will explain how this property can be exploited to speed up the inference of evolutionary parameters by aggregating multiple generations into one. Meanwhile, we note that the two different specifications of the fitness function $g(x;s)$ can be mapped onto each other via the relation $s=\ln(1+s_0)$. For small values of $s$ (or $s_0$) we further have that $s\approx s_0$.

\subsection{Beta-with-Spikes approximation}

Allele frequency data sampled at different time points may not be one generation apart. In this case, it is necessary to sum (\ref{eq:WFTransitionProb}) over multiple intermediate generations to obtain the appropriate DAF.  This is not viable in practice for large population size $N$ and large numbers of intermediate generations $k$, as the memory requirements for this procedure scale as $O(N^2)$, and its computational complexity as $O(kN^3)$ \citep{ref:BwSComparisonTechnicalParis}. These considerations motivate approximating the DAF with some distribution that has a small number of parameters, and using (\ref{eq:WFTransitionProb}) to determine how those parameters change over multiple generations.

\cite{ref:BWS1} introduced the Beta-with-Spikes (BwS) distribution for this purpose. Starting at generation $t$ with a fixed frequency $x_t$, the distribution after $k$ generations is assumed to be well-described by the form
\begin{multline}
\label{eq:BWS}
    {\rm Pr}_{\rm BwS}(x_{t+k}|x_t,N,s) = P_{0,k}\delta(x_{t+k})+P_{1,k}\delta(1-x_{t+k})\\
    {}+\left(1-P_{1,k}-P_{0,k}\right)\frac{x_{t+k}^{\alpha_k-1}(1-x_{t+k})^{\beta_k-1}}{\text{B}(\alpha_k,\beta_k)} \, ,
\end{multline}
which has four parameters $\alpha_k$, $\beta_k$, $P_{0,k}$ and $P_{1,k}$. The central part of this distribution is a Beta distribution, whose shape is controlled by $\alpha_k$ and $\beta_k$ and whose normalization is given by the Beta function ${\rm B}(\alpha_k,\beta_k)$. This distribution has been employed by itself as an approximation to the Wright-Fisher DAF \citep{ref:Hui2015}, and is well-adapted for that purpose by virtue of being defined on the interval $0\le x\le1$ (unlike the Gaussian distribution which assigns a finite probability to unattainable frequencies). Furthermore, a variety of shapes can be accessed by tuning $\alpha_k$ and $\beta_k$, including a uniform distribution, distributions that are strongly peaked around the mean, and those that have an integrable divergence at the boundaries.

This flexibility is however insufficient to capture the accumulation of probability at the boundary points which occurs when an allele goes to fixation. This possibility is incorporated into (\ref{eq:BWS}) through the two Dirac delta function contributions at the extremes of the interval. The quantities $P_{0,k}$ and $P_{1,k}$ then correspond to the probability that alleles $a$ and $A$ have fixed after $k$ generations, respectively. The Beta distribution contribution then describes the DAF \emph{conditioned} on fixation not yet having occurred.

The crucial step in applying the BwS approximation to data is to estimate the parameters $\alpha_k$, $\beta_k$, $P_{0,k}$ and $P_{1,k}$. The general approach is to estimate moments of the DAF and fixation probabilities in the Wright-Fisher model, and choose the parameters in the BwS distribution so that they match up.  To this end, we note first that by defining $a_k$ and $b_k$ as
\begin{align}\label{eq:alphabeta}
    \alpha_k &=\left(\frac{E_k^*(1-E_k^*)}{V_k^*}-1\right)E_k^*\\
    \beta_k  &=\left(\frac{E_k^*(1-E_k^*)}{V_k^*}-1\right)(1-E_k^*) \;,
\end{align}
the mean and variance of the Beta part of the BwS approximation will match the mean $E_k^*$ and variance $V_k^*$ of the Wright-Fisher DAF after $k$ generations, conditioned on fixation not having occurred. These can be obtained from the mean $E_k$ and variance $V_k$ of the full distribution, as well as the fixation probabilities $P_{0,k}$ and $P_{1,k}$, via
\begin{align}
    E_k^*&=\frac{E_k-P_{1,k}}{1-P_{0,k}-P_{1,k}}\\
    V_k^*  &=\frac{V_k+E_k^2-P_{1,k}}{1-P_{0,k}-P_{1,k}}-(E_k^*)^2 \;.
\end{align}
It remains to estimate $E_k$, $V_k$ and the two fixation probabilities.

\subsection{Truncated Taylor-series estimation scheme}

Previous approaches to estimation \citep{ref:Tataru2,ref:BwSComparisonTechnicalParis} have been based around a Taylor-series expansion of the fitness function. Since we will use results from this method in a comparison below, we briefly summarize the procedure.

By recursively applying the law of total probability to the Wright-Fisher transition probability (\ref{eq:WFTransitionProb}), the DAF after $k+1$ generations with starting frequency $x_t$ takes the form:
\begin{equation}\label{eq:TransitionPExact}
    {\rm Pr}_{\rm WF}(x_{t+k+1}|x_t)=\sum_{x_{t+k}} \text{Pr}_{\text{WF}}(x_{t+k+1}|x_{t+k})\text{Pr}_{\text{WF}}(x_{t+k}|x_t)\,
\end{equation}
As set out in Section S1 of the Supplementary Methods, one can use this expression to derive exact recursions for the mean, variance, loss probability and fixation probability. These are
\begin{align}
    E_{k+1} =&\mathbb{E}\left[g(x_k)\right] \\
    V_{k+1} =&\left(1-\frac{1}{N}\right)\text{Var}\left[g(x_k)\right]\nonumber \\
    &+\frac{1}{N}\mathbb{E}\left[g(x_k)\right]\left(1-\mathbb{E}\left[g(x_k)\right]\right) \\
    P_{0,k+1}= &\mathbb{E}\left[\left(1-g(x_k)\right)^N\right]\\
    P_{1,k+1}= &\mathbb{E}\left[g(x_k)^N\right]
\end{align}
where the mean and variance on the right-hand side are with respect to the Wright-Fisher DAF after generation $k$. These recursions are closed in the first two moments only for linear fitness functions $g(x)$, and therefore cannot be computed exactly for nonzero selection coefficients. Closure can be obtained by Taylor expanding $g(x_k)$ about $E_k$ to second order and dropping any higher-order moments that appear (see Supplementary Methods S1.1). This yields
\begin{align}\label{eq:EBadapprox}
    E_{k+1} &\approx g(E_k)+\frac{1}{2}V_k \, g''(E_k) \\
    V_{k+1} &\approx \frac{1}{N}E_{k+1}\left(1-E_{k+1}\right)+\left(1-\frac{1}{N}\right)V_k \, g'(E_k)^2 \;.
\end{align}
This, however, is not exact, with the error at each step increasing with the selection coefficient $s$. Consequently, this recursion is expected only to be valid for small $s$.  The fixation probabilities can be estimated by considering the probability at each boundary within the Beta part of the distribution \citep{ref:Tataru2}
\begin{align}
    P_{0,k+1}&\approx P_{0,k}+(1-P_{0,k}-P_{1,k})\frac{B(\alpha_k,\beta_k+N)}{B(\alpha_k,\beta_k)} \\
\label{eq:P1Badapprox}
    P_{1,k+1}&\approx P_{1,k}+(1-P_{0,k}-P_{1,k})\frac{B(\alpha_k+N,\beta_k)}{B(\alpha_k,\beta_k)}
\end{align}
Despite the approximations made, these recursion relations benefit from being simple and quick to apply.

\subsection{Self-contained estimation scheme}
\label{sec:selfcontained}

Here we take a different approach to estimating the BwS parameters, which is motivated by the expectation that it will keep the accumulation of error under control. The basic idea is to take the BwS distribution obtained after $k$ generations, and generate the intermediate distribution
\begin{equation}\label{eq:TransitionPApprox}
    {\rm Pr}_{\rm int}(x_{t+k+1}|x_t)=\int \text{Pr}_{\text{WF}}(x_{t+k+1}|x_{t+k})\text{Pr}_{\text{BwS}}(x_{t+k}|x_t) {\rm d}x_{t+k}\,
\end{equation}
by applying just one step of the Wright-Fisher process (\ref{eq:WFTransitionProb}). We then examine the moments and fixation probabilities of this intermediate distribution, and use the values obtained to set the BwS parameters for generation $k+1$. We view this as a self-contained estimate, as it maps directly from one set of BwS parameters to the next. 

\begin{figure}[t]
\centering
\includegraphics[width=\linewidth]{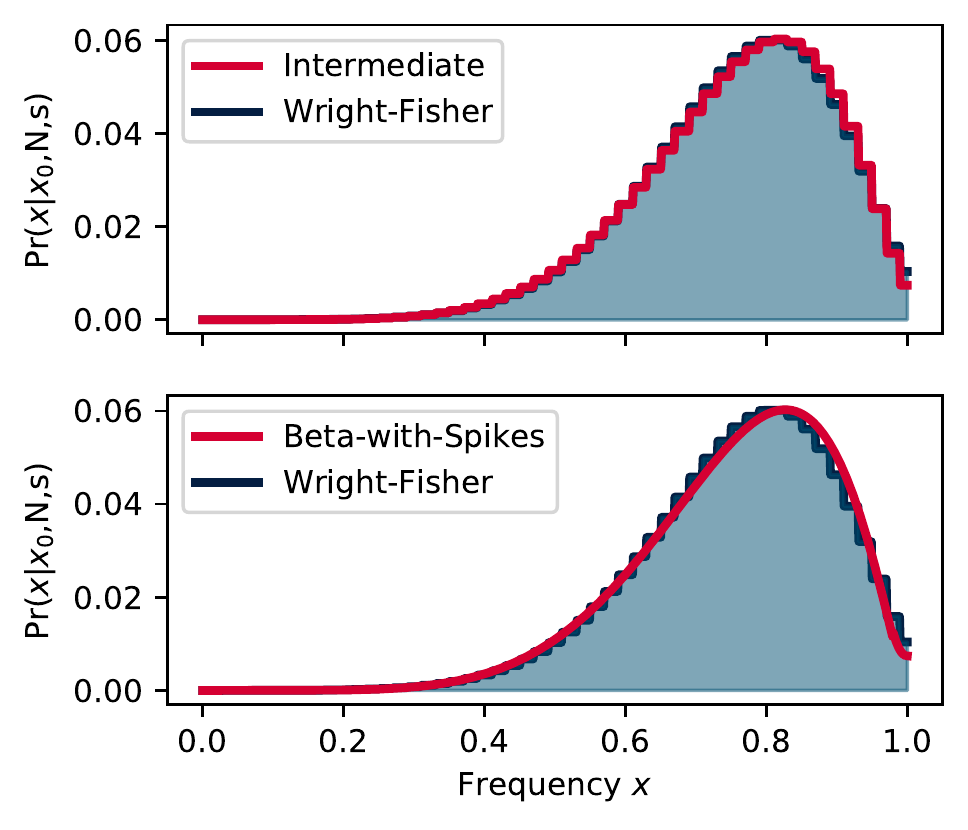}
\caption{Top panel: Comparison of the intermediate (red) and Wright-Fisher (blue) distributions. Bottom panel: Comparison of the Beta-with-Spikes (red) and Wright-Fisher (blue) distributions. Distributions generated with $N=50$, $s=0.2$, $x_0=0.5$ after $k=8$ generations.}
\label{fig:IntermediateComparison}
\end{figure}

Figure~\ref{fig:IntermediateComparison} compares the Wright-Fisher transition probability with the intermediate and Beta-with-Spikes distributions for the case of high selection ($s=0.2)$ and high drift ($N=50$). Even in this challenging regime, we find the intermediate and Beta-with-Spikes distributions remain similar to the exact Wright-Fisher transition probability after $k=8$ generations. The continuous Beta-with-Spikes distribution is generated by mapping its mean, variance, loss and fixation probabilities to those of the intermediate distribution. These are all derived in Supplementary Methods Section S1.2.

For the mean and variance, we find
\begin{align}\label{eq:Eapprox}
    E_{k+1} = & P_{1,k}+\left(1-P_{0,k}-P_{1,k}\right) \mathbb{E}_k\left[g(x)\right] \\
    V_{k+1}  = & \left(1-\frac{1}{N}\right)\left[P_{1,k}+\left(1-P_{0,k}-P_{1,k}\right)\mathbb{E}_k \left[g(x)^2\right] \right] \nonumber \\
    & + \frac{1}{N}E_{k+1}-E_{k+1}^2
\end{align}
where $\mathbb{E}_k\left[\cdot\right]$ represents the expectation value under the Beta distribution with parameters $\alpha_k$ and $\beta_k$:
\begin{equation}
    \label{eq:integral}
    \mathbb{E}_k[f(x)]=\frac{\int_0^1 f(x)x^{\alpha_k-1}(1-x)^{\beta_k-1}\text{d}x}{B(\alpha_k,\beta_k)} \;.
\end{equation}
Meanwhile, the loss and fixation probabilities are obtained as
\begin{align}
    P_{0,k+1} &= P_{0,k}+\left(1-P_{0,k}-P_{1,k}\right)\mathbb{E}_k\left[\left(1-g(x)\right)^N\right] \\
\label{eq:P1approx}
    P_{1,k+1} &= P_{1,k}+\left(1-P_{0,k}-P_{1,k}\right) \mathbb{E}_k\left[g(x)^N\right] \;.
\end{align}

This method does involve more computation than the Taylor-series approach, in that four integrals of the type (\ref{eq:integral}) have to be performed in each iteration, which can increase the computation time by a up to a factor of order 10, compared to an iteration of the Taylor-series approach. However, we have found this effort is manageable in practice, and if necessary can be reduced by appealing to scaling properties identified below. In any case, some care is needed to evaluate these numerically when $\alpha_k<1$ or $\beta_k<1$ (or both), as well as when the integrand in Eq.~(\ref{eq:integral}) is sharply peaked around its mode. We set out the numerical integration algorithms in Supplementary Methods Section S2, and furthermore provide a link to the code used to obtain our results in Section \ref{sec:DataAvailability}.

\subsection{Comparison of estimation schemes}

\begin{figure}[t]
\centering
\includegraphics[width=\linewidth]{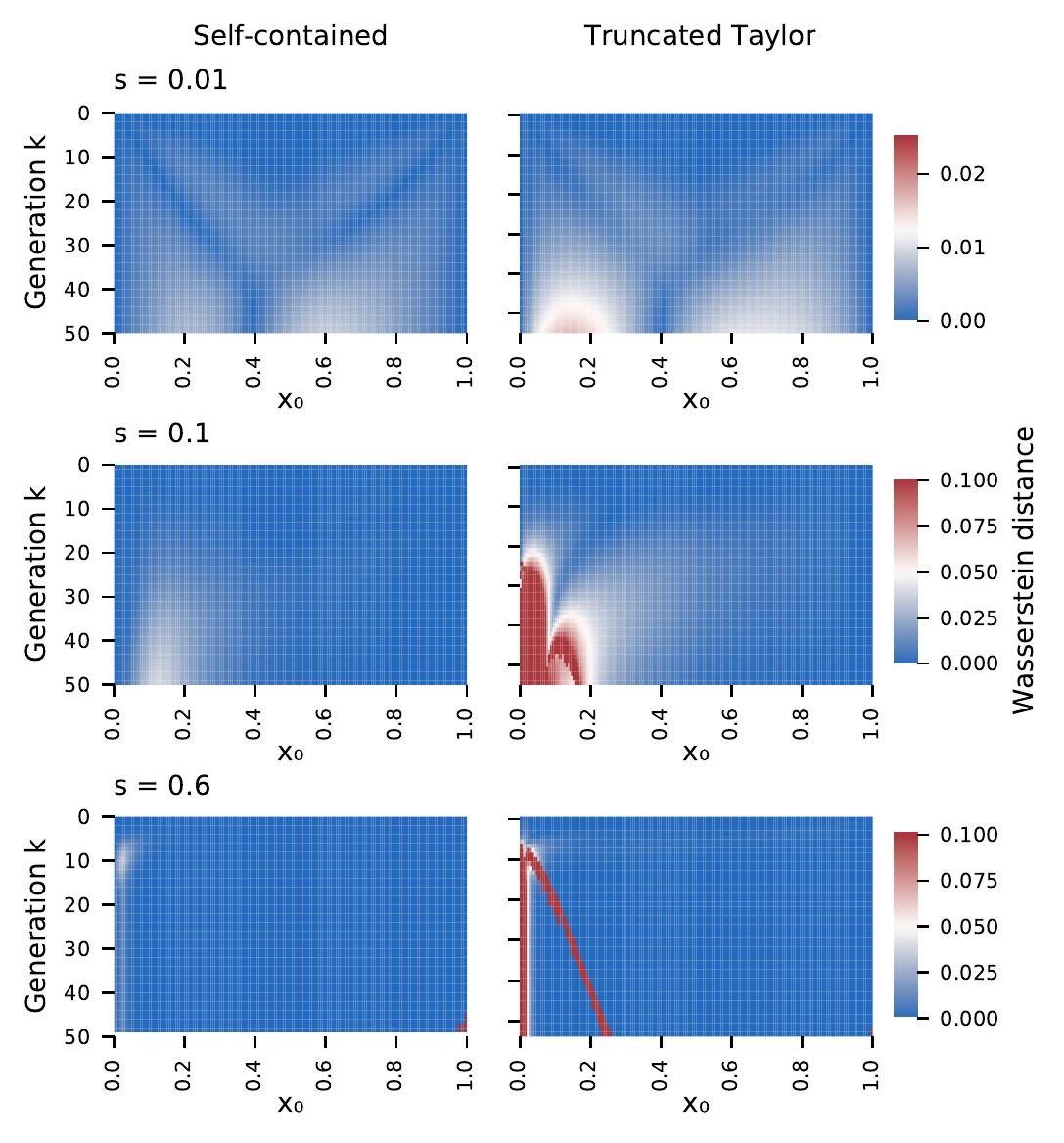}
\caption{Wasserstein distance between the Beta-with-Spikes distribution with numerically exact moments and with approximated moments for two approximation schemes and three values of the selection strength $s$, as a function of the initial frequency $x_0$ and the generation $k$. Left: results for the self-contained approximation. Right: results for the approximation based on the truncated Taylor expansion. Top figures: weak selection ($s=0.01$). Middle figures: intermediate selection ($s=0.1$). Low figures: strong selection ($s=0.6$). Sudden increase of the Wasserstein distance to a value of 1.0 in the approximation based on the truncated Taylor expansion for intermediate and strong selection is due to accumulation of error that leads to an undefined DAF. Results for $N=100$.}
\label{fig:DAFComparison}
\end{figure}

To assess the relative quality of the two estimation schemes, we construct a baseline BwS distribution which is obtained by computing the moments and fixation probabilities within the Wright-Fisher model using numerical methods that are exact to machine precision. Then, we can measure the distance between this baseline and each of the BwS distributions obtained through the estimation schemes set out above. For this purpose, we use the Wasserstein distance, bearing in mind that a smaller distance indicates a better approximation to the baseline. (Note that comparing to the exact Wright-Fisher distribution is difficult, because this is a discrete distribution while the BwS distribution is defined over continuous frequencies.) The results are shown in Figure~\ref{fig:DAFComparison}.

In all cases the population size $N=100$, and we compare performance as a function of the number of generations $k$, the initial frequency $x_0$ and for three different values of the selection coefficient. The estimation based on the truncated Taylor expansion quickly accumulates error in its estimation of the moments when the selection coefficient is sufficiently large, which may lead to negative values of $\alpha$ and $\beta$ and an undefined DAF. This happens at generation $k=23$ for the DAF with intermediate selection and as early as $k=7$ for the DAF with strong selection. This is reflected in the figure as points where the Wasserstein distance reaches a maximal value of 1. The self-contained estimation never leads to an undefined DAF, provided sufficiently accurate integration schemes are used, as $\alpha$ and $\beta$ as defined in Eq.~(\ref{eq:alphabeta}) are always non-negative given that $P_0$, $P_1$, $E$ and $V$ are all obtained from the same well-defined distribution.

\begin{figure}[t]
\centering
\includegraphics[width=\linewidth]{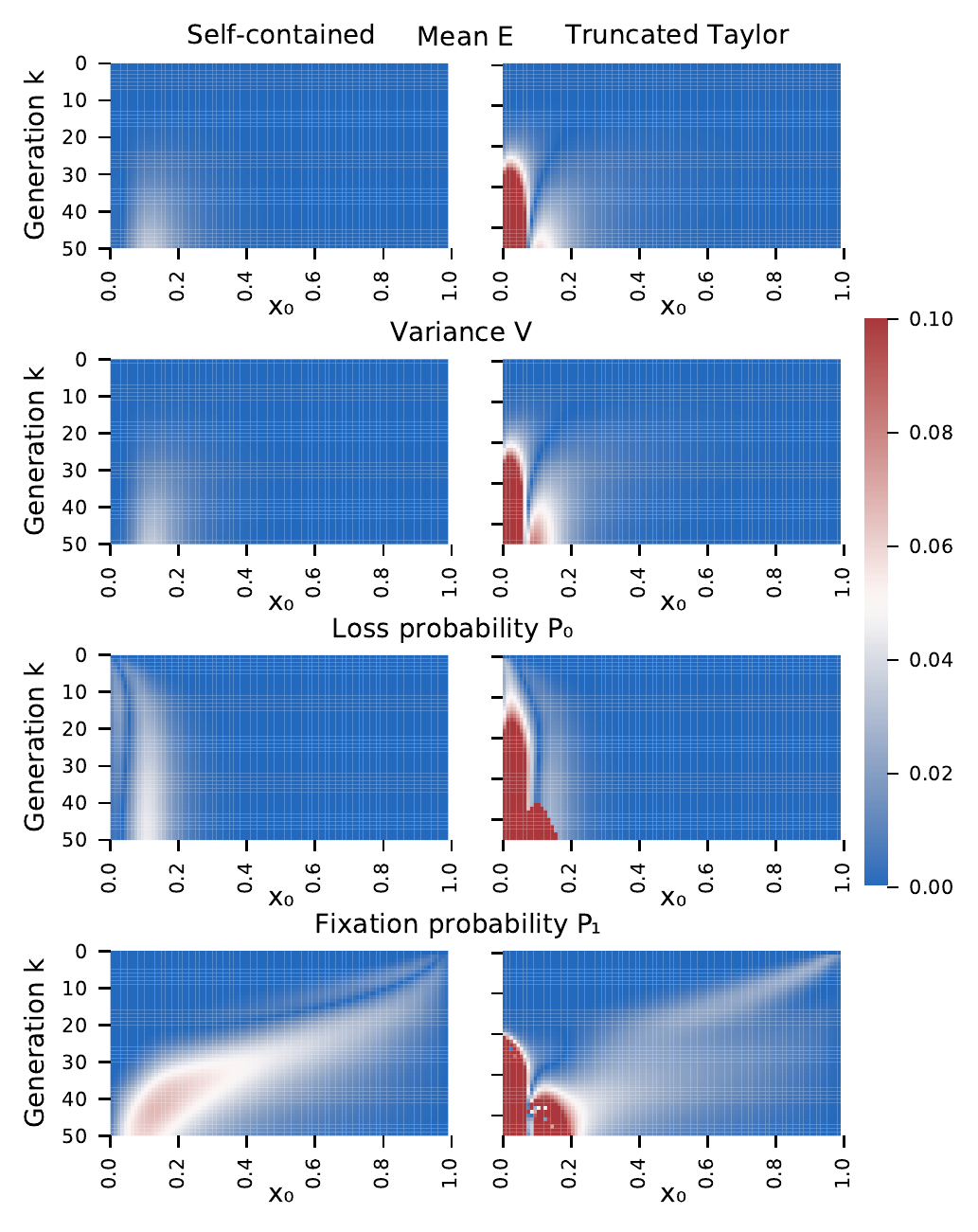}
\caption{Absolute value of the difference between exact and approximated values of the mean, variance, loss probability and fixation probability, as a function of the initial frequency $x_0$ and the generation $k$. Left sub-panels: self-contained approximation. Right sub-panels: truncated Taylor expansion approximation. Results for $N=100$, $s=0.1$.}
\label{fig:MomentComparison}
\end{figure}

Figure \ref{fig:MomentComparison} provides a closer look at the accuracy in the estimation of the mean $E$, variance $V$, loss probability $P_0$ and fixation probability $P_1$ for both the self-contained and truncated Taylor estimation schemes, with $N=100$ and $s=0.1$. For each one of the four parameters, the absolute difference between their estimated and exact values is plotted as a function of the initial probability $x_0$ and generation $k$. Again, the plots demonstrate the robustness of the self-contained estimation, keeping an absolute error below 0.1 for all data points, whereas the truncated Taylor expansion surpasses this value for all parameters before generation $k=30$.

\subsection{Maximum-likelihood inference from time-series data} \label{sec:maxlike}

We now discuss how to infer the effective population size, $N$, and selection coefficient, $s$, through likelihood maximization. The situation we have in mind is when one has samples drawn from the population at a sequence of times $t_1, t_2, \ldots, t_n$. One of the problems we will have to contend with is how many generations of the Wright-Fisher model a particular time interval $t_{i+1}-t_{i}$ corresponds to, which may or may not be known \textit{a priori}. We will set out below a procedure that allows us to deal with this uncertainty, and relate a specific time interval $\Delta t$ to a number of generations $k$.

Another source of uncertainty is in the allele frequencies themselves, as these will be subject to a sampling error that decreases in magnitude as the sample size is increased. In the first instance, we assume that samples are sufficiently large that such error can be neglected. Later, in Section \ref{sec:LanguageApplication}, we set out a scheme that can mitigate against changes in the sample size over time, and will outline extensions to our approach that account for finite sample sizes more rigorously in the Discussion.

Given the above, we can identify the likelihood of the observed data within the BwS approximation as
\begin{equation}\label{eq:accuratesamplinglikelihoodratio}
    L\left(X\,|\,N,s\right)=\prod_i \text{Pr}_{\rm BwS}(x_{t_{i+1}}|x_{t_i},N,s)
\end{equation}
where $X$ is the sequence of frequency measurements $X=(x_{t_1}, x_{t_2}, \ldots, x_{t_n})$. Typically the likelihood function has a single maximum, meaning that the optimal values of $N$ and $s$ can be located straightforwardly with a standard optimization algorithm \citep{ref:Press2007}.

It is often desirable to distinguish neutral from non-neutral evolution, that is, whether the maximum-likelihood value of $s$ is significantly different from zero. This can be achieved by examining the likelihood ratio
\begin{equation}
    \lambda=2\ln\left(\frac{L(X|N,s)}{L(X|N_0,0)}\right)
\end{equation}
in which $N$ and $s$ are the optimal parameters when $s$ is unconstrained, and $N_0$ is the optimal effective population size when the constraint $s=0$ is imposed. The set of all trajectories generated by genetic drift with an effective population size of $N_0$ provide a null distribution for $\lambda$. If, for any empirical trajectory, $\lambda$ is found to lie in the tail of that distribution, we can regard it as significantly different from drift. For long time series, Wilk's theorem \citep{ref:Wilks1938TheLD,ref:CasellaBergerStatisticalInference2001} may hold, and $\lambda$ can be assumed to follow a $\chi^2$ distribution. In the general case, this distribution can be constructed by generating a set of artificial pure drift time series following a Wright-Fisher process with $N=N_0$ and $s=0$. The $p$-value can then be computed as the fraction of these time series whose likelihood-ratio is higher than that of the data of interest, as outlined by \cite{ref:FIT}.

\begin{figure}[t]
\centering
\includegraphics[width=\linewidth]{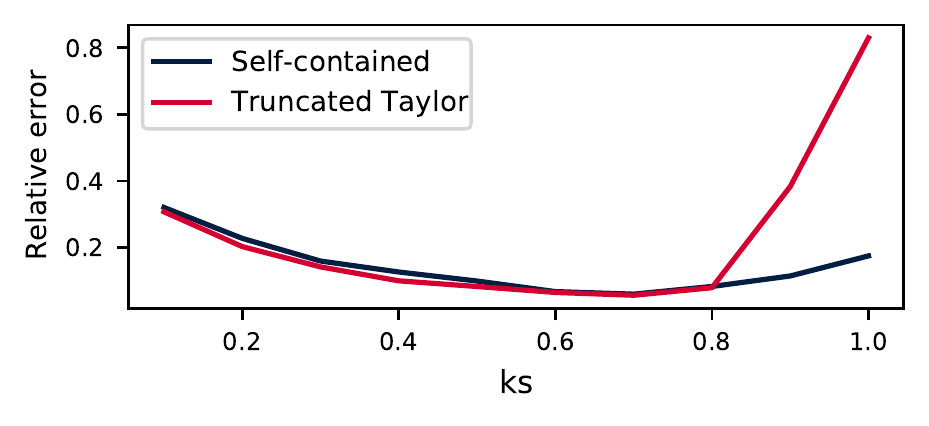}
\caption{Relative error in the estimation of the selection parameter $s$ in artificially generated time series, using a BwS-based maximum-likelihood inference with both the self-contained and the tructated Taylor approximations of the moments, as a function of $ks$, the product of the true selection parameter $s$ and the number of generations $k$ between data points.}
\label{fig:SEstimation}
\end{figure}

Figure \ref{fig:SEstimation} compares the performance of both the self-contained and the truncated Taylor approximations of the moments in the estimation of the selection parameter $s$ from artificially generated time series, using maximum-likelihood inference. While both approximations perform similarly at low $ks$ (the product of the true selection parameter and the number of generations between data points) the numerical stability of the self-contained scheme keeps the error in the estimation decidedly lower starting at $ks=0.8$.

We now turn to the issue of matching a real time interval, $\Delta t$, to a number of generations, $k$, in the Wright-Fisher model. In situations where this is not known, we can appeal to scaling properties of $N$ and $s$ with $k$ to obtain values whose scales are set primarily by $\Delta t$ and only weakly by $k$. As previously mentioned, the fitness function defined by equation~(\ref{eq:s}) satisfies in the deterministic limit ($N\rightarrow\infty$) an exact scaling relation whereby $k$ generations, each lasting $\frac{\Delta t}{k}$, with selection coefficient $s_k$ are equivalent to a single generation, lasting $\Delta t$, with coefficient $s_1=ks_k$. Meanwhile, in the diffusion limit ($N\gg1$) and pure drift ($s=0$), $k$ generations with drift coefficient $N_k$ are equivalent to a single generation with drift coefficient $N_1=N_k/k$ \citep[e.g.][]{ref:WFBook}. In the general case, we propose that for any number of generations $k$ and $\Tilde{k}$ between two data points separated by a time interval $\Delta t$, we have the scaling behaviors
\begin{equation}\label{eq:ScalingBehaviour}
    ks_k=\Tilde{k}s_{\Tilde{k}}\quad N_k/k=N_{\Tilde{k}}/\Tilde{k} \;.
\end{equation}
These relations allow us to divide a time interval into $k$ steps for the purpose of performing the analysis, and quote effective population sizes and selection coefficients appropriately for a standardized time interval determined by $\Tilde{k}$. For example, data could be presented at intervals of ten years, divided into $k=2$ steps of five years for the purposes of analysis, and quoted for a standardized interval of one year ($\Tilde{k}=10$) to facilitate comparison of analyses performed for different time series.

\begin{figure}[t]
\centering
\includegraphics[width=\linewidth]{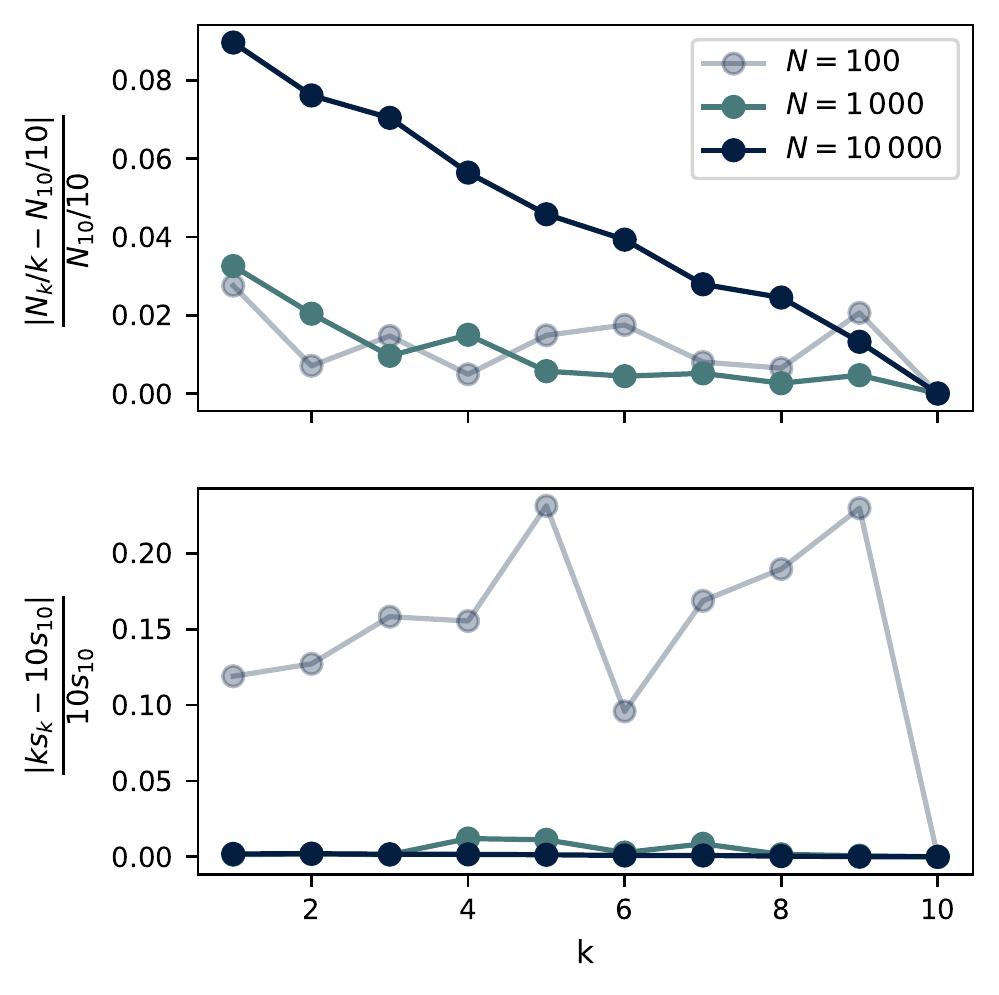}
\caption{Relative error in the scaling behaviour of $N$ and $s$ as time series with 10 generations between data points are reanalysed as having $k<10$ generations between data points. Statistics generated as the average over 2000 artificially-generated time series with $s=0.05$ and $N=100$, $N=1000$, $N=10000$.}
\label{fig:ScalingBehaviour}
\end{figure}

The success of this approach depends on (\ref{eq:ScalingBehaviour}) holding with reasonable accuracy for general $N$ and $s$, beyond the special limits described above. Fig.~\ref{fig:ScalingBehaviour} confirms this for the case of synthetic data generated by iterating (\ref{eq:WFTransitionProb}) ten times between updates. In this case, the true number of generations between sample points is $\Tilde{k}=10$, but we choose to analyze with a different number, $k$. The error in the maximum likelihood estimates of $N_k/k$ and $ks_k$, relative to the true values, is shown as a function of $k$ in Fig.~\ref{fig:ScalingBehaviour}. We find that the error to be modest (around $10\%$ or less) even when the parameters are far from the values that make the scaling behaviour exact (low $N$ for the scaling behaviour of $s$, high selection strength $Ns$ for the scaling behaviour of $N$). What this means in practice is that one can reduce the number of iterations of the self-contained estimation scheme (section \ref{sec:selfcontained}) to a small number $k$, by dividing the time between data points into $k$ generations, whilst retaining reasonable parameter estimates for some chosen standardized generation time.

\subsection{Time-dependent evolutionary parameters} \label{sec:MLEVariable}

Many previous works on analysing evolutionary time series have assumed constant effective population size and selection coefficient. We may however be interested in situations where these parameters change. For example, in cultural evolution, the selection coefficient may change due to one form of behavior gaining social prestige or being stigmatized. Similarly, in genetic evolution, the appearance of a new predator or pathogen could affect an organism's fitness. 

It is straightforward to extend the maximum-likelihood approach outlined above to the case where different parameters apply over different time intervals.  In particular, a single abrupt change can be modeled as two sets of parameters $(N,s)$ that apply before and after a transition time $T$. For $t<T$, the parameters take values $(N_1,s_1)$, and for $t>T$, they take values $(N_2,s_2)$. The optimal parameters $N_1$, $s_1$, $N_2$, $s_2$ and $T$ for a frequency time series $X$ can be found by maximizing the likelihood function $L(X\,|\,N_1,s_1,N_2,s_2,T)$ with respect to all five parameters.

Again, one can use likelihood ratios to determine whether the time division provides a significantly better explanation of the data. Specifically we consider the ratio
\begin{equation}\label{eq:LRTransitionTime}
    \lambda=2\ln\left(\frac{L(X\,|\,N_1,s_1,N_2,s_2,T)}{L(X\,|\,N,s)}\right)
\end{equation}
which compares the optimal likelihood of a model where both $N$ and $s$ change at a time $T$ with one where $N$ and $s$ are fixed for the entire trajectory. Since these models are not nested (on account of the simpler model being found by setting the division point $T$ to its maximum or minimum possible value), $\lambda$ cannot be assumed to be $\chi^2$-distributed. Instead, a $p$-value must be obtained by constructing the empirical distribution of $\lambda$ from trajectories in which there is no time division. This approach can be extended to multiple abrupt changes by further subdividing the trajectories.

\subsection{Sensitivity to changes in selection strength}

To test the ability of this algorithm to detect changes in selection strength, we generate artificial time series by iteratively sampling allele frequencies from Eq.~(\ref{eq:WFTransitionProb}) for $T$ generations. At generation $T/2$, the selection strength goes from $s=0$ to $s=\Delta s$. For each set of values of $500\leq N\leq 10000$, $6\leq T\leq 50$ and $0.001\leq\Delta s\leq 0.3$, we generate 2000 time series, compute their likelihood ratios using Eq.~(\ref{eq:LRTransitionTime}) and find the associated $p$-values. If a $p$-value is under the standard significance threshold, $p<0.05$, the change in selection is considered detected.

\begin{figure}[t]
\centering
\includegraphics[width=\linewidth]{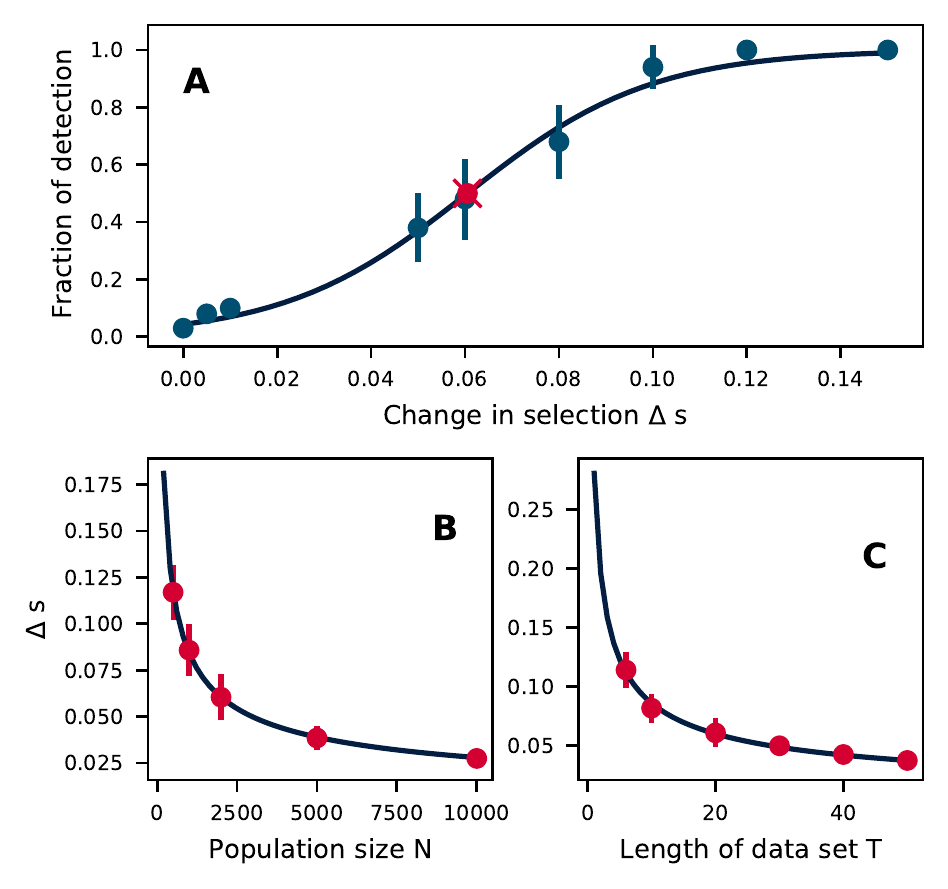}
\caption{(A) Fraction of significant selection as a function of $\Delta s$ for $N=2000$, $T=20$, together with fitted logistic function and estimated characteristic value of $\Delta s$. The fitted logistic function (Eq.~\ref{eq:logistic}) has parameters $a=-3.1\pm0.3$, $b=51\pm5$, $r^2=0.992$. (B) Characteristic $\Delta s$ as a function of $N$ for fixed $T=20$. Power law has proportionality constant $c=0.284\pm0.015$, exponent $d=-0.52\pm0.02$, and $r^2=0.994$. (C) Characteristic $\Delta s$ as a function of $T$ for fixed $N=2000$. Power law has parameters $c=2.33\pm0.12$, $d=-0.481\pm0.008$, $r^2=0.999$. Blue dots represent empirical points obtained as the average of the detection of change in 2000 artificially generated time series. Red dots represent characteristic values of $\Delta s$, obtained from interpolated logistic functions.}
\label{fig:CP}
\end{figure}

Figure \ref{fig:CP}a shows the fraction of time series for which changes in $s$ are detected as a function of the change in selection strength $\Delta s$ at fixed $N=2000$ and $T=20$. As expected, this fraction grows monotonically with $\Delta s$. Empirically, we find that it is well fit by the logistic function
\begin{equation} \label{eq:logistic}
    f(\Delta s)=\frac{1}{1+\exp\left(-a-b\Delta s\right)}
\end{equation}
which allows us to identify a characteristic $\Delta s$ through the value for which the fraction of detected changes equals one half. That is above this value, we are more likely to detect the change than not.

Figures~\ref{fig:CP}b and \ref{fig:CP}c show how this characteristic value varies with effective population size $N$ and the number of generations $T$, respectively. We find that the larger the effective population size or the longer the time series, the smaller the change that can be detected. In both cases this is expected: fluctuations diminish as $N$ increases, thereby increasing the signal-to-noise ratio. Similarly, longer time series provide more information and allow stronger inferences to be drawn. Fig.~\ref{fig:CP}b indicates that relatively small changes are detectable within trajectories of a modest length (e.g., around $T=10$ generations).

\section{Applications to empirical data}

Having validated the methods of the previous section with synthetic data, we now apply them to empirical data from previously published studies. In doing so, we demonstrate their applicability to both genetic and cultural evolution. 

\subsection{Beneficial mutations in yeast populations} \label{sec:BioApplication}

\begin{figure}[t]
\centering
\includegraphics[width=\linewidth]{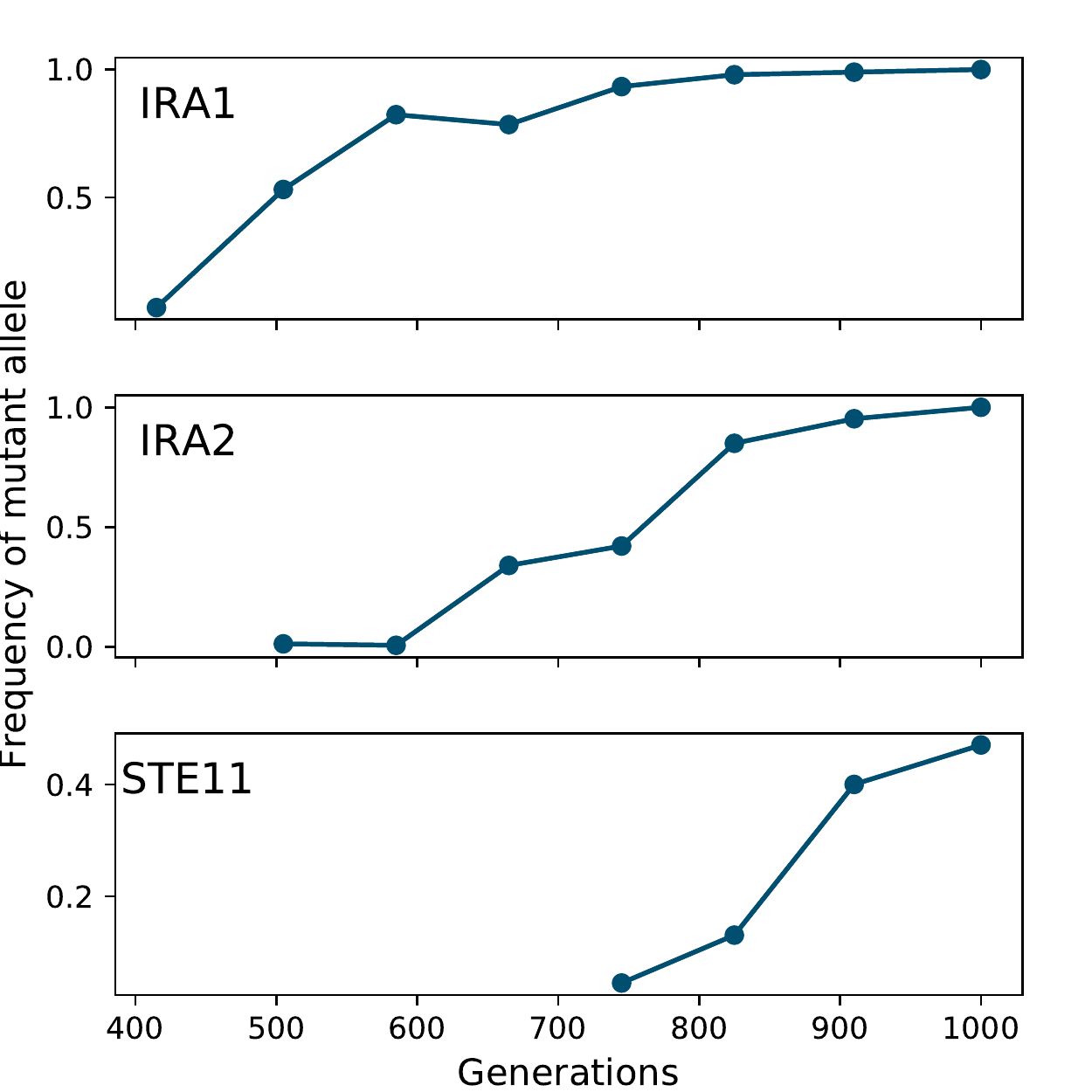}
\caption{Left: trajectory of allele-frequencies of mutation D579Y in gene STE11 in population RMB2-F01. Centre: trajectory of allele-frequencies of mutation Y822* in gene IRA1 in population RMS1-D12. Right: trajectory of allele-frequencies of mutation A2698T in gene IRA2 in population BYS2-D06}
\label{fig:genes}
\end{figure}

We first analyze data from an experiment carried out by \cite{ref:lang2011genetic}, in which 592 populations of baker's yeast (\textit{Saccharomyces cerevisiae}) were evolved over 1000 generations in a rich environment. Several of these populations were deep sequenced, revealing the presence of many adaptive mutations \citep{ref:lang2013pervasive}. \cite{ref:FIT} identified three mutations, affecting genes STE11, IRA1, and IRA2, which were likely to be beneficial, this based on their appearance and spreading in several populations \citep{ref:lang2013pervasive}. The mutant allele frequency trajectories are shown in Figure~\ref{fig:genes}. To test for selection, \cite{ref:FIT} used two methods, each based on a Gaussian approximation to the DAF. The first of these uses the distribution of the likelihood ratio, as described in section \ref{sec:maxlike}. The second is a simpler test, based around the idea that rescaled differences between allele frequencies at subsequent time points should all be drawn from a standard Normal distribution. Despite the independent evidence that all three mutations were being selected for, and in spite of applying the methods to arbitrarily selected subsets of the time series, only one of the six analyses (the likelihood ratio test applied to the last four data points in the IRA1 time series) showed a significant $p$-value for selection \citep[][Table S2]{ref:FIT}.

Here we repeat the analysis using the likelihood ratio test combined with the BwS approximation and both the self-contained and the truncated Taylor estimation schemes. One advantage of methods based on the BwS approximation is that there is no need to truncate the time-series to exclude fixation events, a step that was required in the analysis of \cite{ref:FIT}. There is also no incentive to exclude problematic points close to boundary values because the BwS approximation stays robust. We use a time unit of $\Tilde{k}=5$ generations (the greatest common divisor of the number of generations between data points for all three time series), and rescale $s$ and $N$ as described under Section \ref{sec:maxlike}.

Our results using the self-contained approximation of the moments on the untruncated time series are shown in Table~\ref{tab:genes}, indicating that we find a significant $p$-value in two cases, consistent with the independent evidence of selection. The analysis of gene STE11 produces a low, albeit not below the threshold of significance, $p$-value of $0.084$. This additional sensitivity to selection compared to previous analyses most likely derives from improved handling of the DAF when allele frequencies approach the boundary values. Note how models with selection have a greater optimal population size $N$ than those that rely solely on drift to explain the behavior of the data, as greater effective population size corresponds to lower drift and more deterministic trajectories.

Our results using the truncated Taylor-series scheme, shown in Table~\ref{tab:genesTAYLOR}, show qualitatively similar results, but higher $p$-values for all three time series. This is likely due to the numerical instability of the truncated Taylor method, which produces diverging likelihood-ratios in the empirical statistics used to compute the $p$-value, artificially increasing its value. These observations, combined with the results for synthetic data, suggest that the self-contained estimation scheme allows selection to be more reliably detected. It is also of note that the computation time for these frequency time series under the self-contained scheme was only 7 times slower than its truncated Taylor counterpart, perfectly manageable for this type of analysis.

\begin{table}[b]
\caption{\label{tab:genes}%
Results of the analysis of the allele-frequency trajectories of genes STE11, IRA1 and IRA2 using the self-contained scheme for the estimation of the moments. $N_0$: optimal population parameter under the null-model of pure drift. $N$: optimal population parameter under model with selection. $s$: optimal selection strength.
}
\begin{ruledtabular}
\begin{tabular}{ccccc}
\textrm{Gene}& $N_0$ & $N$ & $s$ & \textbf{$p$-value} \\
\colrule
STE11 & 1350  & 1860 &	0.010 & 0.084 \\
 IRA1  & 1240 & 1970 & 0.0091 & 0.006  \\
 IRA2  & 987   & 1980 &	0.019 & 0 \\
\end{tabular}
\end{ruledtabular}
\end{table}

\begin{table}[b]
\caption{\label{tab:genesTAYLOR}%
Results of the analysis of the allele-frequency trajectories of genes STE11, IRA1 and IRA2 using the truncated Taylor-series scheme for the estimation of the moments. $N_0$: optimal population parameter under the null-model of pure drift. $N$: optimal population parameter under model with selection. $s$: optimal selection strength.
}
\begin{ruledtabular}
\begin{tabular}{ccccc}
\textrm{Gene}& $N_0$ & $N$ & $s$ & \textbf{$p$-value} \\
\colrule
 STE11 & 627  & 1590 &	0.0097 & 0.19 \\
 IRA1  & 526 & 1440 & 0.015 & 0.014  \\
 IRA2  & 377   & 794 &	0.018 & 0.022 \\
\end{tabular}
\end{ruledtabular}
\end{table}

\subsection{Regulated and unregulated language change} \label{sec:LanguageApplication}

\begin{figure}[t]
\centering
\includegraphics[width=\linewidth]{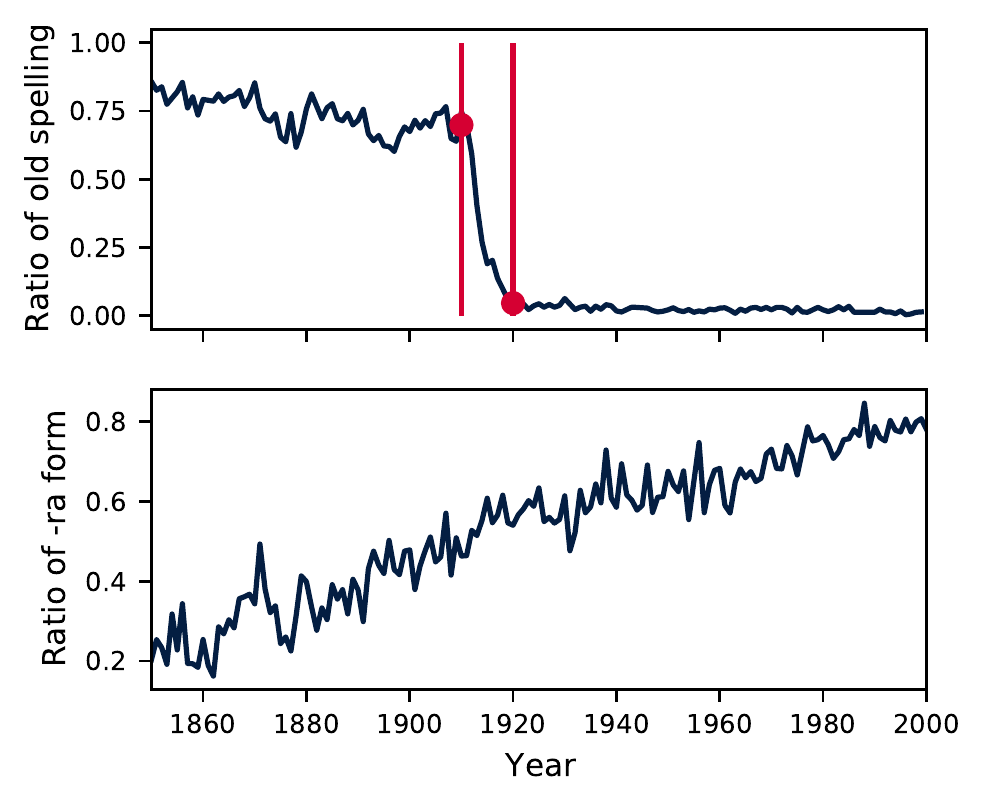}
\caption{Top: regulated change. Trajectory of frequency of usage in the Spanish Google Books corpus of old, accentuated spellings of words \textit{a}, \textit{e}, \textit{o} and \textit{u}, with detected years of change in the selection parameter marked with red vertical lines. The first detected year in 1910 is only a year before the true year of introduction the orthographic reform by the RAE which declared the old spellings non-standard. Bottom: unregulated change. Trajectory of frequency of usage in the Spanish Google Books corpus of the -ra- form of the past subjunctive, as opposed to the -se- form, for which the time-divided model is not significant.}
\label{fig:culture}
\end{figure}

We now turn to cultural evolution and examine the dynamics of historical changes occurring in the 2019 update to the Spanish Google Books corpus \citep{ref:GoogleBooks}. We look at both a regulated and unregulated change that occurred between the 19th and 20th centuries \citep{ref:DynamicsofNormChangeCulturalApplicationAmato}. The regulated change was a spelling reform introduced by the Real Academia Española, the central regulatory institution of the Spanish language, in their \textit{Gramática de la lengua castellana} \citep{ref:RAEGramatica1911}, and entailed the accentuated \textit{á} (meaning \textit{to}), \textit{ó} (meaning \textit{or}), \textit{é} (alternative form of \textit{y}, meaning \textit{and}), and \textit{ú} (alternative form of \textit{ó}) being replaced by their unaccented forms \textit{a}, \textit{o}, \textit{e} and \textit{u}. The unregulated change occurred in the absence of such an intervention, and involved the competition dynamics between two completely equivalent forms of the past subjunctive tense, with verbal affixes -ra- and -se-. Thus, the third person singular of the past subjunctive of the verb \textit{evolucionar} (to evolve) could be either \textit{evolucionara} or \textit{evolucionase}. Both forms of the past subjunctive are considered completely equivalent in all contexts. In spite of this, in the last 150 years, there has been a steady transition in the corpus, from a clear preference of the -se- form to a clear preference of the -ra- form (see lower panel of Figure~\ref{fig:culture}). As of yet, there is no agreed upon explanation of this phenomenon, from either corpus-based or sociolinguistic perspectives \citep{GuzmanNaranjoRaSeCorpus,KempasRaSeSociolinguistics}

The processes of regulated and unregulated change have previously been modeled by the cultural analogs of mutation and migration in a large population \citep{ref:DynamicsofNormChangeCulturalApplicationAmato}. In the notation of the present work, this corresponds to a linear fitness function $g(x)=ax+b$ and a large fixed value of the effective population size $N$, with residuals modeled by a standard Normal distribution rather than the Wright-Fisher model. Given that we are dealing with a competition between pre-existing variants, we consider it more natural to view the evolution as being driven by selection, albeit where the selection coefficient may change over time, for example, due to the imposition of the reform, or because social preferences and norms can change over time. To this end, we turn to the method described in Section \ref{sec:MLEVariable} to detect changes in the selection coefficient with fluctuations in variant frequencies accounted for through the cultural analog of genetic drift (whose amplitude may also change over time).

An issue that we have to contend with when dealing with historical language data is that the sample sizes change over time. Specifically, the general trend is for data to become scarcer as earlier time periods are examined. The increased sampling fluctuations at early times could then be misattributed to drift, that is, the intrinstic fluctuations in the cultural transmission process, rather than the sampling of linguistic data from the population.  One way to address this is to create subsamples of the larger data sets in the time series, with the subsample size chosen in such a way that the contribution from sampling is of equal magnitude across the time series. Then, any detected change in the effective population size must be due to changes in the intrinsic fluctuations, rather than sampling. In practice, we achieve this by generating for each time point a binomial random variable with a success probability equal to that of the original sample, but a sample size $m$ given by
\begin{equation}
m = \frac{m_0}{1-\frac{m_0}{n}} \;.
\end{equation}
where $n$ is the original sample size, and $m_0$ is the smallest original sample size across the entire time series. This formula is derived in Supplementary Methods Section S3.

After constructing the resampled time series, we apply the method of time-dependent evolutionary parameters (section \ref{sec:MLEVariable}) to estimate parameter and $p$-values were found for models with and without a single time division. When this time-divided model has a $p$-value below $0.05$, we repeat the process for each of the subseries, accepting subsequent time divisions whenever $p<0.05$.  The result of this analysis is shown in Figure~\ref{fig:culture}, with estimated parameter and $p$-values given in  Tables~\ref{tab:VariableModelUnregulated} and \ref{tab:VariableModelRegulated}.

A single time division model is not found to be significant for the process of an unregulated change. Thus suggests that it has not been driven by any abrupt change in the social perception of either the -ra- or -se- forms of the past subjunctive. By contrast, we find that a first division at $T=1910$ followed by a second division at $T=1920$ are both significant in the case of a regulated change. The first time division delimits an early period where the accented spellings (\textit{á}, \textit{ó}, \textit{é}, \textit{ú}) were widely used, and one where they rapidly fell out of use. The division point falls at the start of the decline, and is in fact only one year before the introduction of the reform \citep{ref:RAEGramatica1911}. Thus it seems likely that the reform caused individual language users to change their attitude towards the accented forms. We also note that the estimated effective population size does not change across this first time division. The second time division falls at the end of the period of decline, and we note from Table~\ref{tab:VariableModelRegulated} that the selection coefficient is estimated to be much smaller than during the transition period. It is perhaps the case that modern Spanish speakers encounter the accented forms sufficiently rarely that they do not hold any particular disposition towards it. The significance or otherwise of the change in effective population size is somewhat less clear, and we do not speculate further. We conclude this Section by noting that if one does not account for the possibility of the selection coefficient changing in time, one does not find a significant effect of selection (according to the likelihood ratio test discussed in section \ref{sec:maxlike}). 

\begin{table}[b]
\caption{\label{tab:VariableModelUnregulated}%
Results for the analysis of unregulated change in the affixes of past subjunctive verbal forms in Spanish between the years 1850 and 2000, using time-divided models. The time-division is found to not be significant.
}
\begin{ruledtabular}
\begin{tabular}{cccccc}
 $T$ & $N_1$ & $N_2$ & $s_1$ & $s_2$ & $p$  \\
\colrule
1961 & 46.9 & 124 & 0.016 & 0.026 & 0.15
\end{tabular}
\end{ruledtabular}
\end{table}

\begin{table}[tb]
\caption{\label{tab:VariableModelRegulated}%
Results for the analysis of regulated change in the accentuation of single-letter words in Spanish between the years 1850 and 2000, using time-divided models. Two time divisions are found to be significant, in 1910 and 1920, delimiting the transition process between the old and new spelling rules introduced by the RAE in 1911.
}
\begin{ruledtabular}
\begin{tabular}{cccccccccc}
 $T_1$ & $T_2$ & $N_1$ & $N_2$ & $N_3$ & $s_1$ & $s_2$ & $s_3$ & $p_1$ & $p_2$ \\
\colrule
 1910 & 1920 & 63 & 72 & 197 & -0.002 & -0.45 & -0.013 & 0.0 & 0.0
\end{tabular}
\end{ruledtabular}
\end{table}



\section{Discussion}\label{sec:disco}

In this work, we have introduced a method for obtaining reliable maximum-likelihood estimates of parameters within the Wright-Fisher model from time-series data for allele frequencies. Our approach is underpinned by the Beta-with-Spikes (BwS) distribution \citep{ref:BWS1,ref:Tataru2} which, despite not exactly matching the distribution of allele frequencies within the Wright-Fisher or related models, captures its essential features. These are the possibility of extinction or fixation of an allele, accounted for by the spikes, and that unfixed alleles are governed by a continuous distribution that is well-characterized by its mean and variance.

The challenge in utilising the BwS approximation is accurately determining appropriate parameter values. Earlier works \citep{ref:Tataru2,ref:BwSComparisonTechnicalParis} used Taylor series expansions to estimate how parameter values should change from one generation to the next. These have the benefit of being simple to evaluate, but the truncation of the Taylor series results in the approximation being unreliable when the selection coefficient is large. In particular it can generate parameter values that cause the BwS distribution to be ill-defined.

Here, we have turned to a \emph{self-contained} approximation, where the BwS distribution is used as the initial condition for one step of Wright-Fisher evolution, and a fresh BwS distribution is fit to the intermediate distribution that results. In this approach, the approximating distribution remains well-defined, and provides an adequate approximation to the Wright-Fisher model even when selection is strong. We have demonstrated the reliability of the method for the Wright-Fisher model with frequency-independent selection by comparing distributions directly, and by determining the error on the maximum-likelihood estimate of the selection coefficient for artificial time series where the true value is known.

The method is however more computationally intensive than the Taylor-series approach, since it is necessary to compute four integrals over a BwS distribution at each generation. Nevertheless, we find that the method can be applied to data for both genetic and cultural evolution without undue computational effort. In particular, we can reduce the number of calculations that need to be performed by aggregating multiple generations into a single effective generation, and appeal to the scaling properties of the effective population size and the selection strength when this is done, as described in Section \ref{sec:maxlike}.

In Section \ref{sec:BioApplication}, we found that we were able to obtain a significant signal of selection for two out of three genes for which there is independent evidence of selection \citep{ref:lang2013pervasive,ref:FIT}, whereas other methods either break down or do not yield a uniformly significant result, even when time series are truncated. Although we cannot be certain that the gene frequencies were driven by selection in all cases, our results suggest that the inability to reject the null hypothesis of drift in \cite{ref:FIT} may lie in the sensitivity of the test that was applied. 

As noted in the introduction, cultural evolutionary processes, such as language change, can also be couched in evolutionary terms \citep{ref:Cavalli,ref:BoydRicherson,ref:CroftLanguageChange} and furthermore represented mathematically by the Wright-Fisher model \citep{ref:UtteranceSelection,ref:WordsAsAlleles,ref:RichardPLOS}. Until recently, the analysis of historical corpus data within this framework has been hampered by the limited availability of methods that can be applied to variant frequency time-series data. In a pioneering work, \cite{ref:FITApplication} applied the method of \cite{ref:FIT} to assess the relative contributions from drift and selection in historical changes, and found that drift was a likely explanation in many cases. However, this analysis suffers from the same potential lack of sensitivity as was seen in the application to genetic data \citep{ref:FIT}.

Furthermore, our approach lends itself to extensions that allow for the possibility that evolutionary parameters may change over time. In Section \ref{sec:LanguageApplication} we demonstrated this in the context of regulated and unregulated change, showing that changes in selection strength that might reasonably be expected in regulated change are detected by our method, whereas no such changes were found in the the case of unregulated change.

A limitation of the method we have employed here is the assumption that sample sizes are large enough that the uncertainty on allele frequency estimates drawn from them can be neglected. In reality, this may not be the case, under which circumstances one would normally turn to a hidden Markov framework, as proposed by \cite{ref:bollback2008estimation} in the context of genetic time series data. A na\"{\i}ve numerical implementation of this scheme would involve integrating over each of the (now hidden) frequencies $x_{t_i}$ in Eq.~(\ref{eq:accuratesamplinglikelihoodratio}), which dramatically increases the computational demands. One way to circumvent the additional integrals is to employ an expectation-maximisation algorithm. However, we have found that if the effective population size is considered a free parameter, expectation-maximisation tends to push this towards infinity due to piecewise deterministic trajectories being favored by the algorithm. Therefore, some further work is needed to develop tractable methods for jointly estimating effective population size and the selection coefficient when working with data subject to sampling uncertainty.

In the meantime, we have shown how one can account for known variation in sample sizes over the course of the time series, which is important when trying to determine if the effective population size (which governs fluctuations intrinsic to the evolutionary process) changes over time. The basic idea is to reduce the size of the larger samples so that the uncertainty due to sampling is then uniform across the time series. Although the resulting estimates of the effective population size then contain a contribution from both intrinsic fluctuations and sampling, any detected changes in the effective population size are most likely to arise from a change in the amplitude of the intrinsic fluctuations.

In summary, despite certain limitations, the method introduced here allows evolutionary parameters to be reliably estimated, and when combined with empirical likelihood ratio tests, can be used to test departure from a variety of null hypotheses. Although we have focused on the Wright-Fisher model with frequency-independent selection, it could be extended to models that involve other evolutionary processes. Extensions to processes involving more than two alleles are likely also possible in principle, although may involve higher-dimensional integrals that become difficult to perform numerically.

\section{Data availability}
\label{sec:DataAvailability}

The code, as well as data used in Section \ref{sec:LanguageApplication} are available at 

\url{https://datashare.ed.ac.uk/handle/10283/4811}

Data used in Section \ref{sec:BioApplication}, originally obtained by \cite{ref:lang2011genetic}, is available in the supplementary materials of \cite{ref:FIT}.

\section{Funding}

JGM holds Principal's Career Development Scholarship awarded by the University of Edinburgh.

For the purpose of open access, the authors have applied a Creative Commons Attribution (CC BY) licence to any Author Accepted Manuscript version arising from this submission.

\section{Conflicts of interest}

The authors declare no conflicts of interest.

\bibliography{Arxiv_Final}
\newpage
\
\newpage
\appendix

\section{Derivation of the estimation schemes}

The estimation schemes each start with exact recursions for the mean and variance of the DAF at generation $t+k+1$. Given a transition probability ${\rm P}(x_{t+k}|x_t)$ representing the DAF after $k$ time steps with starting frequency $x_t$, the DAF at the next timestep can be found using the Chapman-Kolmogorov equation for the transition probabilities of the process:
\begin{equation}\label{eq:Intermediate}
    {\rm P}(x_{t+k+1}|x_t)=\int\text{Pr}_{\text{WF}}(x_{t+k+1}|x_{t+k}){\rm P}(x_{t+k}|x_t){\rm d}x_{t+k} \\
\end{equation}
which arises from the law of total probability. In it, the one-step Wright-Fisher transition probability with population size $N$ is given by
\begin{equation}\label{eq:WFTransitionProb}
    {\rm Pr}_{\rm WF}(x'|x)=
    \binom{N}{Nx'}
    g\left(x\right)^{Nx'}\left(1-g\left(x\right)\right)^{N(1-x')} \;.
\end{equation}

From this, the mean of the frequency at generation $t+k+1$ is given by
\begin{align}\label{eq:MeanP}
    E_{k+1} &= \sum_{n=0}^N\frac{n}{N}{\rm P}\left(x_{t+k+1}=\frac{n}{N}|x_t\right)\nonumber \\
    &=\sum_{n=0}^N\frac{n}{N} \int\text{Pr}_{\text{WF}}\left(\frac{n}{N}|x_{t+k}\right)\text{P}\left(x_{t+k}|x_t\right){\rm d}x_{t+k}\nonumber \\
    &= \int\sum_{n=0}^N\frac{n}{N} \text{Pr}_{\text{WF}}\left(\frac{n}{N}|x_{t+k}\right)\text{P}\left(x_{t+k}|x_t\right){\rm d}x_{t+k}\nonumber \\
    &= \int g\left(x_{t+k}\right)\text{P}\left(x_{t+k}|x_t\right){\rm d}x_{t+k}\nonumber\\
    &= \mathbb{E}_{\text{P}}\left[g(x_{t+k})\right]\, ,
\end{align}
where the third equality uses the analytical result for the mean of the Wright-Fisher transition probability, thus eliminating the explicit dependence of the final result on this probability function. In the last equality, $\mathbb{E}_{\text{P}}$ represents the expectation under the transition probability $\text{P}$.

The variance, similarly, can be found as
\footnotesize
\begin{align}\label{eq:VarianceP}
    V_{k+1} =& \sum_{n=0}^N\left(\frac{n}{N}-E_{k+1}\right)^2{\rm P}\left(x_{t+k+1}=\frac{n}{N}|x_t\right)\nonumber \\
    =& \int \sum_{n=0}^N\left(\frac{n}{N} \right)^2\text{Pr}_{\text{WF}}\left(\frac{n}{N}|x_{t+k}\right)\text{P}(x_{t+k}|x_t) {\rm d}x_{t+k} -E_{k+1}^2\nonumber\\
    =& \int \left[\left(1-\frac{1}{N}\right)g(x_{t+k})^2+\frac{1}{N}g(x_{t+k})\right]\text{P}(x_{t+k}|x_t) {\rm d}x_{t+k}-E_{k+1}^2\nonumber \\
    =& \left(1-\frac{1}{N}\right)\left(\mathbb{E}_{\text{P}}[g(x_{t+k})^2]-E_{k+1}^2\right) +\frac{1}{N}\left(\mathbb{E}_{\text{P}}[g(x_{t+k})]-E_{k+1}^2\right)\nonumber \\
    =&  \left(1-\frac{1}{N}\right)\text{Var}_{\text{P}}[g(x_{t+k})] +\frac{1}{N}\mathbb{E}_{\text{P}}[g(x_{t+k})]\left(1-\mathbb{E}_{\text{P}}[g(x_{t+k})]\right)
\end{align}
\normalsize
where the third equality uses the analytical form of the second moment of the Wright-Fisher transition probability, and the last equality uses equation \ref{eq:MeanP} together with the definition of the variance. In it, $\text{Var}_{\text{P}}$ represents the variance under the transition probability P.

The loss probability is given by
\begin{align}\label{eq:LossP}
    P_{0,k+1} =& {\rm P}\left(x_{t+k+1}=0|x_t\right)\nonumber \\
    =& \int \text{Pr}_{\text{WF}}\left(0|x_{t+k}\right)\text{P}(x_{t+k}|x_t) {\rm d}x_{t+k}\nonumber \\
    =& \int \left(1-g(x_{t+k})\right)^N\text{P}(x_{t+k}|x_t) {\rm d}x_{t+k}\nonumber \\
    =& \mathbb{E}_{\text{P}}\left[\left(1-g(x_{t+k})\right)^N\right]
\end{align}
and equivalently for the fixation probability we have
\begin{align}\label{eq:FixationP}
    P_{1,k+1} =& {\rm P}\left(x_{t+k+1}=1|x_t\right)\nonumber \\
    =& \int \text{Pr}_{\text{WF}}\left(1|x_{t+k}\right)\text{P}(x_{t+k}|x_t) {\rm d}x_{t+k}\nonumber \\
    =& \int g(x_{t+k})^N\text{P}(x_{t+k}|x_t) {\rm d}x_{t+k}\nonumber \\
    =& \mathbb{E}_{\text{P}}\left[g(x_{t+k})^N\right]
\end{align}

\subsection{Derivation of the truncated Taylor scheme}

Equations \ref{eq:MeanP} to \ref{eq:FixationP} do not have closed analytical forms for arbitrary $g(x_{k+t})$, and their direct integration using a $k$-generation Wright-Fisher transition probability is computationally intractable in general. From equations \ref{eq:MeanP} to \ref{eq:FixationP}, recursive relations for the moments, loss and fixation probabilities after $k+1$ generations can be obtained by Taylor expanding $g(x_{t+k})$ about $E_k$ up to second order, and dropping all moments of higher order than the variance.

For the mean (from equation \ref{eq:MeanP}):
\begin{align}\label{eq:MeanT}
    E_{k+1}\approx&g(E_k)+g'(E_k)\mathbb{E}_{\text{P}}\left[\left(x_{t+k}-E_k\right)\right]\nonumber \\
    &+\frac{g''(E_k)}{2}\mathbb{E}_{\text{P}}\left[\left(x_{t+k}-E_k\right)^2\right]\nonumber\\
    =&g(E_k)+\frac{g''(E_k)}{2}V_k
\end{align}
where we have used the identities $\mathbb{E}_{\text{P}}\left[x_{t+k}\right]=E_k$ and $\mathbb{E}_{\text{P}}\left[\left(x_{t+k}-E_k\right)^2\right]=V_k$.

For the variance:
\small
\begin{align}
    \text{Var}_{\text{P}}[g(x_{t+k})] =& \mathbb{E}_{\text{P}}\left[g(x_{t+k})^2\right]-E_{k+1}^2 \nonumber\\
    \approx&g(E_k)^2-E_{k+1}^2+2g'(E_k)g(E_k)\mathbb{E}_{\text{P}}\left[(x_{t+k}-E_k)\right] \nonumber\\
    &+\left(g'(E_k)^2+g''(E_k)g(E_k)\right)\mathbb{E}_{\text{P}}\left[(x_{t+k}-E_k)^2\right]\nonumber\\
    =&-\left(\frac{g''(E_k)}{2}V_k\right)^2-g''(E_k)g(E_k)V_k \nonumber\\
    & +\left(g'(E_k)^2+g''(E_k)g(E_k)\right)V_k \nonumber\\
    =& -\left(\frac{g''(E_k)}{2}V_k\right)^2+g'(E_k)^2V_k \nonumber \\
    \approx& g'(E_k)^2V_k
\end{align}
\normalsize
where, in line with previous work (see supplementary material in \cite{ref:BwSComparisonTechnicalParis}), the last equality assumes $V_k^2$ to be of the same order as $\mathbb{E}_{\text{P}}\left[\left(x_{t+k}-E_k\right)^4\right]$ and thus negligible. By introducing this result into equation \ref{eq:VarianceP}, we obtain
\begin{equation}
    V_{k+1} \approx \frac{1}{N}E_{k+1}\left(1-E_{k+1}\right)+\left(1-\frac{1}{N}\right)V_k \, g'(E_k)^2\, .
\end{equation}

For the loss and fixation probabilities, \cite{ref:Tataru2} propose linearizing the fitness function as $g(x)=x$ and taking the transition probability $P(x_{t+k}|x_t)$ to be a Beta-with-Spikes:
\begin{multline}
\label{eq:BWS}
    {\rm Pr}_{\rm BwS}(x_{t+k}|x_t) = P_{0,k}\delta(x_{t+k})+P_{1,k}\delta(1-x_{t+k})\\
    {}+\left(1-P_{1,k}-P_{0,k}\right)\frac{x_{t+k}^{\alpha_k-1}(1-x_{t+k})^{\beta_k-1}}{\text{B}(\alpha_k,\beta_k)} \, ,
\end{multline}

With that, starting from equation \ref{eq:LossP}:
\small
\begin{align}
    P_{0,k+1} \approx& \mathbb{E}_{\text{BwS}}\left[\left(1-x_{t+k}\right)^N\right] \nonumber \\
    =& P_{0,k}+(1-P_{0,k}-P_{1,k})\int\frac{x^{\alpha_k+1}(1-x)^{\beta_k+N+1}{\rm d}x}{B(\alpha_k,\beta_k)} \nonumber \\
    =& P_{0,k}+(1-P_{0,k}-P_{1,k})\frac{B(\alpha_k,\beta_k+N)}{B(\alpha_k,\beta_k)}
\end{align}
\normalsize
and similarly for the fixation probability (equation \ref{eq:FixationP}):
\begin{equation}
    P_{1,k+1}\approx P_{1,k}+(1-P_{0,k}-P_{1,k})\frac{B(\alpha_k+N,\beta_k)}{B(\alpha_k,\beta_k)}\, .
\end{equation}

\subsection{Derivation of the self-contained scheme}

We take the intermediate distribution at time $t+k$ to be given by
\small
\begin{equation}\label{eq:Intermediate}
    {\rm Pr}_{\rm int}(x_{t+k+1}|x_t)=\int \text{Pr}_{\text{WF}}(x_{t+k+1}|x_{t+k})\text{Pr}_{\text{BwS}}(x_{t+k}|x_t) {\rm d}x_{t+k}\, .
\end{equation}
\normalsize

With this, the integrals in equations \ref{eq:MeanP} to \ref{eq:FixationP} are computationally tractable without having to resort to approximations of the fitness function.

From equation \ref{eq:MeanP}, the mean of the intermediate distribution with population size $N$ is given by
\begin{align}\label{eq:MeanS}
    E_{k+1} &= \mathbb{E}_{\text{BwS}}\left[g(x_k)\right]\nonumber \\
    &= P_{1,k}+\left(1-P_{0,k}-P_{1,k}\right) \frac{\int g(x)x^{\alpha_k-1}(1-x)^{\beta_k-1}}{\text{B}(\alpha_k,\beta_k)}\, ,
\end{align}
where the last equality uses equation \ref{eq:BWS} together with the assumptions $g(0)=0$ and $g(1)=1$.

The variance, similarly, can be found as
\small
\begin{align}\label{eq:VarianceS}
    V_{k+1} =& \left(1-\frac{1}{N}\right)\left(\mathbb{E}_{\text{BwS}}[g(x_{t+k})^2]-\mathbb{E}_{\text{BwS}}[g(x_{t+k})]^2\right) \nonumber \\ 
    &+\frac{1}{N}\mathbb{E}_{\text{BwS}}[g(x_{t+k})]\left(1-\mathbb{E}_{\text{BwS}}[g(x_{t+k})]\right) \nonumber \\
    =& \left(1-\frac{1}{N}\right)\left[P_{1,k}+\left(1-P_{0,k}-P_{1,k}\right) \frac{\int g(x)^2x^{\alpha_k-1}(1-x)^{\beta_k-1}}{\text{B}(\alpha_k,\beta_k)}\right]\nonumber \\
    & +\frac{1}{N}E_{k+1}-E_{k+1}^2\, ,
\end{align}
\normalsize
where equations \ref{eq:VarianceP}, \ref{eq:BWS} and \ref{eq:MeanS} have been used.

The loss probability (equation \ref{eq:LossP}) is given by
\small
\begin{align}\label{eq:LossS}
    P_{0,k+1} =& \mathbb{E}_{\text{BwS}}\left[\left(1-g(x_{t+k})\right)^N\right] \nonumber \\
    =& P_{0,k}+\left(1-P_{0,k}-P_{1,k}\right) \frac{\int \left(1-g(x)\right)^Nx^{\alpha_k-1}(1-x)^{\beta_k-1}}{\text{B}(\alpha_k,\beta_k)}
\end{align}
\normalsize
and equivalently for the fixation probability (equation \ref{eq:FixationP}) we have
\small
\begin{align}\label{eq:FixationS}
    P_{1,k+1} =& \mathbb{E}_{\text{P}}\left[g(x_{t+k})^N\right]\nonumber \\
    =& P_{1,k}+\left(1-P_{0,k}-P_{1,k}\right) \frac{\int g(x)^Nx^{\alpha_k-1}(1-x)^{\beta_k-1}}{\text{B}(\alpha_k,\beta_k)}
\end{align}
\normalsize

These parameters may be used now to generate the parameters $\alpha_{k+1}$ and $\beta_{k+1}$ of the Beta-with-Spikes transition probability after $k+1$ generations.

\section{Numerical integration techniques}

To implement the self-contained Beta-with-Spikes approximation, we are required to evaluate integrals of the general form
\begin{equation}
\label{int}
    I=\int_0^1\frac{x^{\alpha-1}(1-x)^{\beta-1}}{(1+(e^s-1)x)^\gamma}\text{d}x \;.
\end{equation}
Specialized libraries may be able to deal with these. For people wishing to implement them in programming languages that do not have these libraries available, some care is needed in their computation. We set out these details below.

Depending on the values of $\alpha$ and $\beta$, the integrand may diverge at either endpoint, or be sharply peaked at some point $0<x<1$. Special handling is needed around these points. 

\medskip

More precisely, when $\alpha<1$, the integrand diverges as $x\to0$. In this situation, we split the range of integration at $x=\Delta x$, and expand the integral over $0<x<\Delta x$ in powers of $\Delta x$ to second order. We find
\small
\begin{multline}
    \int_0^{\Delta x} \frac{x^{\alpha-1}(1-x)^{\beta-1}}{(1+(e^s-1)x)^\gamma}\text{d}x  \approx\frac{1}{\alpha}\Delta x^\alpha-\frac{\gamma S+\beta-1}{\alpha+1}\Delta x^{\alpha+1} \\ 
    {}+\frac{(\beta-1)\gamma S+\frac{1}{2}(\beta-1)(\beta-2)+\frac{1}{2}S^2\gamma(\gamma+1)}{\alpha+2}\Delta x^{\alpha+2}
\end{multline}
\normalsize
where $S=e^s-1$.
Similarly, when $\beta<1$, there is a divergence as $x\to1$ which can be handled by splitting the integral at $x=1-\Delta x$. The corresponding expansion is
\small
\begin{multline}
    \int_{1-\Delta x}^{1} \frac{x^{\alpha-1}(1-x)^{\beta-1}}{(1+(e^s-1)x)^\gamma}\text{d}x  \approx\frac{1}{\beta e^{-\gamma s}}\Delta x^\beta-\frac{\gamma \tilde{S}+\alpha-1}{(\beta+1)e^{-\gamma s}}\Delta x^{\beta+1} \\ 
    {} +\frac{(\alpha-1)\gamma \tilde{S}+\frac{1}{2}(\alpha-1)(\alpha-2)+\frac{1}{2}\tilde{S}^2\gamma(\gamma+1)}{(\beta+2)e^{-\gamma s}}\Delta x^{\beta+2}
\end{multline}
\normalsize
where $\tilde{S}=1-e^{-s}$. The integral is split at both boundaries when $\alpha<1$ and $\beta<1$.

The value of $\Delta x$ is obtained numerically as the distance from the boundary at which the integrand reaches a fixed, high value (1000 in our implementation) to ensure $\Delta x$ is small enough to make the previous Taylor expansions accurate, as this ensures $\Delta x \ll \frac{1}{1000}$. The rest of the integral is computed using trapezoid rule with adaptive quadrature \citep{ref:Press2007}.

\medskip

When $\alpha>1$, $\beta>1$ and $s\ne0$, the derivative of the integrand has four roots, located at $x=0$, $x=1$ and
\begin{equation}
    x_{\pm}=\frac{A\pm\sqrt{A^2+4(\alpha-1)B}}{2B}
\end{equation}
where
\begin{align}
    A &=(\alpha-1)e^s-(\beta-1)-(e^s-1)\gamma \\
    B &=(e^s-1)(\alpha-1+\beta-1-\gamma)
\end{align}
Since the integrand is always $0$ at $x=0$ and $x=1$ and is positive, continuous and differentiable in between these two values, it must have and odd number of maxima in the $(0,1)$ interval. However, since there are only two stationary points in this interval, the integrand has only a single maximum, located at either $x_+$ or $x_-$.

To increase accuracy when this maximum is strongly peaked, the integral is split into two at the maximum, each half being computed using trapezoid rule with adaptive quadrature. This assures that the maximum is not missed by the adaptive rule in scenarios where it is narrower than the initial step, and slightly increases the speed of convergence of the integration method.

\medskip

Finally, when $\alpha>1, \beta>1$ and $s=0$, there is a maximum at
\begin{equation}
    x_0=\frac{\alpha-1}{\alpha+\beta-2} \;.
\end{equation}
This can be handled in the same way as in the $s\ne0$ case.

\section{Sampling error equalisation}

When dealing with time-series data, estimates of an allele's (or cultural variant's) frequency may derive from samples of different sizes at different times, and therefore be subject to greater or lesser degrees of sampling error. To disentangle this from fluctuations in the underlying frequencies themselves (arising, for example, from genetic drift), it is helpful to equalize the amplitude of the sampling error across the time series. Then, any changes in the amplitude of the resulting fluctuations over time can be ascribed to the process that generates the underlying frequencies, as changes in how samples are constructed have already been accounted for.

In the main text, we describe a resampling procedure that effects this equalization. Given a sample of size $n$ at some time $t$, within which a fraction $x$ of items are of one particular variant (i.e., a specific allele or word form), we construct a binomial sample of size $m$ and success probability $x$ in such a way that the variance of the corresponding variant frequency $y$ is consistent with being derived from a binomial sample of fixed size $m_0$. The key point to note here is that the original sampling process already contributes some variance to $y$. Therefore, $m$ will depend on the original sample size so that the additional variance arising from resampling gives the desired overall variance.

To determine the appropriate sample size $m$, we consider the first two moments of the random variable $y$.  Given some value of $x$ via the original sampling process, the binomial resampling process implies that
\begin{align}
\mathbb{E}(y|x) &= x \\
\mathbb{E}(y^2|x) &= \left(1-\frac{1}{m}\right) x^2 + \frac{1}{m} x \;.
\end{align}
We now average over all possible realizations of the original sampling process to determine the first two moments of the resampled frequency $y$, finding
\begin{align}
\label{Ey}
\mathbb{E}(y) &= \mathbb{E}(x) \\
\label{Eyy}
\mathbb{E}(y^2) &= \left(1-\frac{1}{m}\right) \mathbb{E}(x^2) + \frac{1}{m} \mathbb{E}(x) \;.
\end{align}
Although the true variant frequency $p$ is unknown, we have that
\begin{align}
\mathbb{E}(x) &= p \\
\mathbb{E}(x^2) &= \left(1-\frac{1}{n}\right) p^2 + \frac{1}{n} p 
\end{align}
where $n$ is the original sample size. Substituting these expressions into (\ref{Ey}) and (\ref{Eyy}), we find that
\begin{align}
\mathrm{Var}(y) &= \mathbb{E}(y^2) - [\mathbb{E}(y)]^2 \\
&= \left[ 1 - \left(1-\frac{1}{n}\right) \left(1 - \frac{1}{m}\right) \right] p(1-p) \;.
\end{align}
This is the variance that would be obtained if the original sampling process involved a sample of size
\begin{equation}
\frac{1}{m_0} = 1 - \left(1-\frac{1}{n}\right) \left(1 - \frac{1}{m}\right)
\end{equation}
where $m_0$ is the fixed effective sample size introduced above. We note that this result can also be obtained by applying the law of total variance to the pair of random variables $x$ and $y$, where $x$ is drawn from ${\rm Bin}(n,p)$ and $y$ from ${\rm Bin}(m,x)$.

Rearranging, we find that the resampled population size $m$ should be
\begin{equation}
m = \frac{1-\frac{1}{n}}{\frac{1}{m_0}-\frac{1}{n}} \approx \frac{m_0}{1-\frac{m_0}{n}} \;,
\end{equation}
in which the approximation holds when the original sample size $n \gg 1$, which is typically the case. This latter approximate formula is the one that is quoted in the main text.

\medskip

\begin{figure}[t]
\centering
\includegraphics[width=\linewidth]{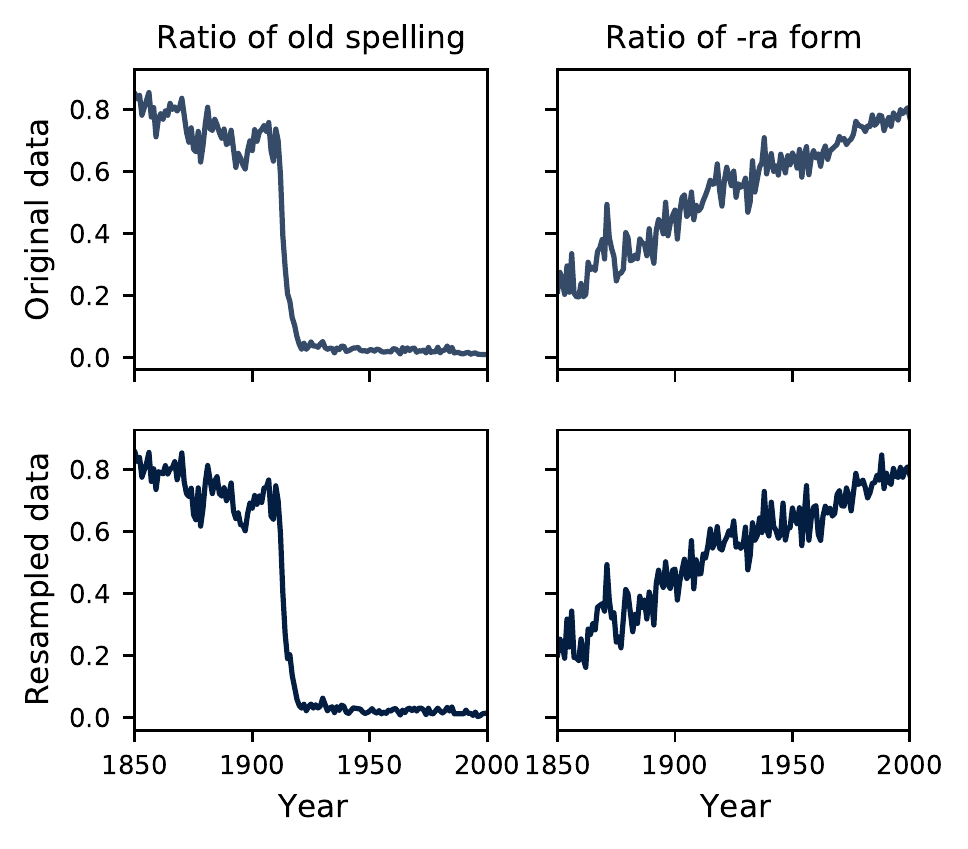}
\caption{Comparison of the Google Books Spanish time series before (upper panels) and after (lower panels) sampling error equalization. Left panels relate to the ratio of old spelling of single-letter words (\textit{a}, \textit{e}, \textit{o}, \textit{u}) and the right panels to the ratio of usage of the \textit{-ra-} form of the past subjunctive, as opposed to the completely equivalent \textit{-se-} form. The effect of the sampling error equalization is particularly evident in the case of the \textit{-ra-} form.}
\label{fig:ResampleComparison}
\end{figure}

The effects of resampling on the Google Books Spanish data can be observed in Figure~\ref{fig:ResampleComparison}. The effects of noise equalization are particularly noticeable in the last 50 years of the ratio of the -ra form data set (right panels). Without applying the equalization (upper right panel), a spurious change in effective population size is detected.  After applying the equalization, we find that the fluctuations maintain a similar amplitude throughout the time series, and no significant change in effective population size is found (as reported in the main text). 

\end{document}